\documentstyle[pre,aps,preprint,psfig]{revtex}
\begin{document}

\title{Using Topological Statistics to Detect 
Determinism in Time Series}

\author{Guillermo J. Ortega}
\address{Centro de Estudios e Investigaciones,
Universidad Nacional de Quilmes,\\ R. S. Pe\~na 180, 1876, Bernal, Argentine}
\author{and}
\author{Enrique Louis}
\address{Departamento de F\'{\i}sica Aplicada,
Facultad de Ciencias,\\
Universidad de Alicante, Apartado 99, E-03020, Alicante, Spain}

\date{\today}

\maketitle

\begin{abstract}
\parbox{14cm}
{Statistical differentiability of the measure along the
reconstructed trajectory is a good candidate to quantify determinism
in time series. 
The procedure is based upon a formula that explicitly shows
the sensitivity of the measure to stochasticity.
Numerical results for partially surrogated time series
and series derived from several stochastic models, illustrate
the usefulness of the method proposed here. The method is shown to
work also for high--dimensional systems and experimental time series. }

\end{abstract}
\vspace*{2.0truecm}

\indent
\section{Introduction}
This paper deals with the problem of determining whether the 
complex behaviour of a single time series may be explained 
in terms of a deterministic or a stochastic dynamical system. 
Although the idea was present
from the very first days of the nonlinear time series analysis \cite {GP83},
methods explicitly aiming to detect determinism in time series 
\cite{KD92,WB93,SC94,SM90} have only been published rather recently.
The fact that noise can mimic or mask deterministic ({\it e.g.} chaotic) 
behavior in classical measures of chaos (Lyapunov exponents,  
$K_2$ entropy and correlation dimension) \cite{OP89,Th91,AB93} has urged
the need for more specific methods to discriminate deterministic
from stochastic behaviour. 

Measures and vector fields \cite{ER85}, densities and trajectories
\cite {LM85}, metric and topological \cite {Gi98}, statistical and geometrical, 
these are, roughly speaking, different terms used
to denominate the two broad approaches for investigating the
chaotic behaviour of dynamical systems. It is a remarkable fact that
most of the aforementioned methods to detect determinism are based in
the study of trajectories or vector fields. 
Considering the noisy character of the inverse problem of time
series analysis (from the point of view of nonlinear dynamics),
the statistical approach must lie at the heart of any proposed
methodology. This becomes relevant as soon as the system underlying a 
time series is 
"excited" beyond the simple bifurcations, so that the geometrical
information about the shape of the attractor, or the motion on it, 
is no longer available. This is also true in the case of stochastic
dynamical system, where noise can make impossible to reconstruct the 
geometrical information \cite{CE91}. Moreover, trajectories
and the geometry of the flow {\it on} the attractor are well 
defined in the case of a deterministic system, in contrast to the
case of a stochastic dynamics.

Bearing that in mind, we have developed a novel approach based
in the study of the differentiability of the natural measure. The
statistical character of our methodology is twofold. The first
one concerns the use of the natural (or physical) measure. Differentiability
of this invariant measure will be considered in the direction of the
evolution, that is, along any typical trajectory. In this way, 
"smoothness" of the trajectory {\it and} measure are considered in
a single step. This is a general result of dissipative systems,
regardless of its application to time series analysis.
The second one is the use of statistics for analyzing mathematical
properties \cite{PC95}, which allow us to test the differentiability
of the measure in a statistical sense. In this way we can extend our
methods to signal analysis.\\
The procedure is based upon a formula that will be derived in the
next section, and  that will allow us to discriminate between
deterministic and stochastic behaviours. This problem was tackled
by us in \cite {OL98} whereas the basic idea of the
methodology was proposed in \cite {Or95,Or96}. 

The paper is organized as follows. Section II is devoted to 
the mathematical background of our approach. At the end of
the section we have included a brief summary of the concepts
introduced in this section. This is addressed to the less inclined 
mathematical reader, which can skip most of the mathematical details given 
in the section.  In Sec. II.A we introduce the fundamentals of our approach. 
The extension to stochastic dynamics is considered in Section II.B.
General concepts about measure projections and time series are
explained in section II.C, whereas Section II.D is devoted
to a brief discussion of the statistics of topological properties 
used in this work.  Section III is devoted to discuss the models to which 
we apply our method and the numerical procedures followed in each case. 
We also propose a variant of the surrogation method, partial
surrogation, as a way to tune the degree of stochasticity.
The results are presented in Section IV. The sensitivity of 
the time derivative of the measure to stochasticity, the dependence of
the results on the embedding dimension and the application of our approach 
to mixed time series are illustrated on the models discussed in Section III. 
We also present a simple test on an experimental time series like
the Belousov-Zhabotinski chemical reaction.

\section {Theoretical Approach}
Here we discuss in detail the mathematical background of our method. The 
natural measure is first defined and the equation of motion that
it obeys explicitly given. Then we show how stochasticity triggers a {\it wild}
behavior of the time derivative of the natural measure. This is
the feature upon which our proposal is based. In order to evaluate 
quantitatively this behavior we borrow the method proposed by Pecora
{\it et al} \cite{PC95}. In particular we evaluate (statistically)
the continuity of the logarithmic derivative of the natural measure. Those
readers more interested in applications may directly go to the Summary at 
the end of the Section where the most important features of the method
are highlighted with the aid of the equations discussed hereafter.

\subsection{The Natural Invariant Measure along the trajectory}
Consider a dissipative dynamical system described by $n$-first-order
differential equations ${\bf {\dot x}} = {\bf F}({\bf x})$.
The corresponding flow ${\it f}^t$ maps a "typical" initial
condition ${\bf x}_0$ into ${\bf x}(t) = {\it f}^t ({\bf x}_{0})$
at time $t$. Once transients are over, the motion settles
over the attractor ${\cal A}$.

Given such a system, we can define a probability space
(measure space) $({\cal A}, {\cal B},\mu)$, where 
${\cal B}$ is the $\sigma$-algebra generated by the open sets of
the invariant set ${\cal A}$, and $\mu$ is an invariant measure defined 
over the sub-sets
of ${\cal B}$ such that $\mu : {\cal B} \rightarrow (0,1)$.

Of all the invariant measures which can be defined,
only one is relevant from the point of view of the experimentalist
or in computer simulations. This is the {\it natural} invariant
measure, which gives the limiting distribution of almost all starting
initial conditions. Using the indicator function
$1_B({\bf y}) = 1$, if ${\bf y} \in B_{\epsilon}$ and $0$ otherwise,
we can define this measure for a set
$B_{\epsilon} ({\bf x}) = \{{\bf y}:d({\bf x},{\bf y) \leq \epsilon}\}$
as,
\begin{eqnarray*}
\mu(B_{\epsilon} ({\bf x}))  = \lim_{t \rightarrow \infty}
{1 \over t} \int_0^t 1_B (f^{\tau}({\bf y}_0)) \: d\tau
\end{eqnarray*}
for almost all ${\bf y}_0$ in the
basin of attraction. The so--defined measure is invariant in the sense
that it remains constant upon application of the evolution
operator $f^{\tau}$,
\begin{eqnarray*}
{\mu}(B_{\epsilon}({\bf x})) = {\mu}(f^{\tau}(B_{\epsilon}({\bf x})) \; .
\end{eqnarray*}

In the present case what we calculate is ${\mu}(B_{\epsilon}({\bf x}(t))$
along the trajectory. This means that the evolution operator changes
the point at which the set $B_{\epsilon}$ is centered, namely,
${\mu}(B_{\epsilon}(f^{\tau}({\bf x})))$, leaving unchanged the ball
$B_{\epsilon}$. This is the so--called
Lagrangian evolution (see Fig. \ref{fig_1}), that has to be distinguished
from the Liouvillian evolution, $\mu(B_{\epsilon}({\bf x}(t)) 
\rightarrow f^{\tau} ({\mu}(B_{\epsilon}({\bf x}(t))$,
which, contracts the volume along the trajectory,
as illustrated in Fig. \ref{fig_1}.

Calling ${\mu}({\bf x}(t)) = \mu(B_{\epsilon}
({\bf x(t)}))$, the "material derivative" of the natural measure can 
be expressed as,
\begin{equation}
\frac{d{\mu ({\bf x}(t))}}{dt}  =
 \dot {\bf x}{\bf .}\nabla \mu
\label{e1}
\end{equation}
\noindent
or
\begin{equation}
\frac{d{\mu ({\bf x}(t))}}{dt} =
{\nabla} {\bf .} (\mu {\bf F}) - \mu
\nabla {\bf .} {\bf F}
\label{e2}
\end{equation}
\noindent

Eq. (\ref{e2}) only involves partial derivatives of
${\mu}({\bf x}(t))$ along the trajectory,
and since each trajectory in the
attractor is contained in the support,
${\mu}({\bf x}(t))$ is smooth along
any trajectory, and is therefore well-defined.
It is known that for axiom-A systems, hyperbolic and with a
dense set of periodic orbits, it is possible to decompose
the tangent bundle at each  ${\bf x} \in {\cal A}$ in two
linear sub-spaces, stable and unstable. 
It is also possible to define a measure in these systems, known as a
SRB (Sinai-Ruelle-Bowen) which have the property of
being smooth along the unstable manifold \cite{ER85},
due fundamentally to the stretching of the trajectories.
On the other side, it is expected to have a wild behavior along the 
stable direction.  In the present case, considering that
Eq. (\ref{e2}) only involves derivatives along the
trajectory and that in this direction no expansion or compression 
of the flux occurs (the associated Lyapunov exponent is equal to zero), 
we expect a smooth behaviour of the measure.

\subsection {Stochastic Dynamics}

Consider now a stochastic dissipative dynamical system
with additive noise, described by $n$-first-order
differential equations,
\begin{eqnarray}
{\bf {\dot x}} = {\bf F}({\bf x}) + \eta {\bf G}(t)
\label{e3}
\end{eqnarray}
where $\eta > 0$ is a small number (noise intensity) and
${\bf G}(t)$ is a vector of independently and identically
distributed random Gaussian variables, of zero mean and correlations
$< G_i(t) G_j (t') > =  \delta_{ij} \delta (t - t')$.
A physical system
will normally have a small level $\eta$ of random noise,
so that it can be considered a stochastic process rather than
a deterministic one. In a computer study, roundoff errors should
play the role of the random noise.
For suitable noise and $\eta$, the stochastic time evolution
(\ref{e3}) has a unique stationary measure $\mu$
\cite{ER85}. This is the {\it natural} (or {\it physical}) invariant measure
defined above.

Inserting Eq. (\ref{e3}) in (\ref{e1}),
\begin{equation}
\frac{d{\mu ({\bf x}(t))}}{dt}  =
 ({\bf F} + \eta {\bf G}(t)). \nabla \mu
\label{e4}
\end{equation}
\noindent Whenever the vector field ${\bf F}({\bf x})$ can be expressed as
${\bf F}({\bf x})=-\nabla \phi({\bf x})+{\bf f}({\bf x})$, with
${\bf f}({\bf x})$ being orthogonal to the gradient term and having
no divergence,  the measure is given by \cite{GT84,ST97}:
\begin{equation}
\mu({\bf x}) = N \exp(- \frac{\phi({\bf x})}{\eta^2})
\label{e5}
\end{equation}
Then, introducing (\ref{e5}) in (\ref{e4}) we arrive at:
\begin{equation}
\frac{d{(\ln \mu ({\bf x}(t)))}}{dt} = \frac{1}{\eta}\left (\frac{1}{\eta}
|{\bf F}({\bf x})|^2 + {\bf G}(t).{\bf F}({\bf x})\right )
\label{e6}
\end{equation}

In section II.A we discussed the smoothness of the measure
along the trajectory for a deterministic system, at least in
the case there exists a SRB measure. In the present case, that
is, a stochastic process, it is possible to define a unique
stationary measure that will tend to the SRB for
$\eta \rightarrow 0$. This is the {\it Kolmogorov
Measure} \cite{ER85}. At each time step of evolution, we add
some noise (of amplitude $\eta$). If we repeat these operations
(evolving by time evolution and putting some noise) again 
and again then we will get an invariant measure which is smooth
along the unstable direction. This is because the deterministic
part of the time evolution will improve the continuity of 
the density in the unstable direction by stretching, and roughening 
it in the other directions due to contraction. Thus, in the zero-noise
limit $\eta \rightarrow 0$ we will get a measure $\mu$ that
satisfies SRB conditions.

Eq. (\ref{e6}) provide an alternative tool to investigate the findings
of \cite{SC94}, according to which smoothness in phase
space implies determinism in time series. For
weak noise levels, the first term of the right hand
side of (\ref{e6}) is dominant over the second term. In
this case, smoothness in phase space implies "continuity"
in the left hand side of (\ref{e6}), or differentiability of
the measure along the trajectory. On the other hand,
in the case of strong noise levels, the second term is
the dominant one and a wild behavior in the
measure must be expected. This is so because
the vector ${\bf G}(t)$ is
uncorrelated with the actual position in the phase space.
Although Eq. (\ref{e6}) was derived for a vector field obeying some
restrictions (see above), our numerical results indicate that it can
be applied to more general systems.

\subsection {Time Series and Measure Projection}

In case that dynamical invariants are to be
estimated from the knowledge of a single observable,
we can rely upon Takens' Delay Coordinate Map Theorem
\cite{Ta81}. According to it, the structure of the attractor
(including differential information) can be
preserved by using delay coordinate maps \cite {PC80},
as long as the embedding dimension $m$ is large enough to
fully unfold the attractor structure.
Furthermore, the Fractal Whitney Embedding Prevalence
Theorem \cite {SY91} tells us that almost every smooth map will
be an embedding (one-to-one and differential structure
preserving) provided that $m > 2D$, where $D$ is the
box-counting dimension of the attractor.
However, as shown in Refs.
\cite{DG93,SS98},
when dealing with functions of the measure, embedding dimensions greater
than the correlation dimension are enough.

From the statistical point of view, we must ask
how is the embedding process in the measure space
$(A,{\cal B})$ associated to the attractor $A$.
That is, if $\mu$ is a measure in $\Re^n$ and
$\phi : \Re^n \rightarrow \Re^m$ is a function, we will
have a projected or {\it induced measure} $\mu_{\pi} = \phi(\mu)$ over
a subset $S \in \Re^m$, where $\phi(\mu)(S) = \mu(\phi^{-1}(S))$.
This is important because we are actually transforming
the measure in the embedding process. 
Most of the existing literature is related with
projections and reconstructions of the invariants sets
and the dimensions of the corresponding probability measures
\cite{Ka68,Ma75,SY97}. In our case,
if $\mu$ is the physical measure describing the original
system ($\mu$ is carried by an attractor in phase space)
then the points in the reconstructed space are
equidistributed with respect to the projected measure
$\mu_{\pi}$ \cite{ER85} (except for particular cases avoided
explicitly by the embedding theorems \cite{SY91}). This is
a sufficient condition which allows us to extract information
about the original system working in its corresponding 
projection.

If we have a time series, we can construct an $m$-dimensional
vector:
\begin{eqnarray*}
\bar {x_i}  = (x_i, x_{i+ \tau},...,x_{i+(m-1).\tau})
\end{eqnarray*}
where $\tau$ is the time delay, chosen by one of the
standard methods \cite{AB93}. Then, from a single set of observations,
multivariate vectors in $m$-dimensional space are used to
trace out the orbits of the system.
Fig. \ref{fig_2} shows the concept of the reconstruction method. 
Explicitly shown is the reconstruction of the measure
carried by the attractor $\mu_{\pi}$.

\noindent

\subsection {Statistics of Topological Properties}

Taken's theorem \cite{Ta81}, and its sequels, give a rigorous 
justification for state space reconstruction. The essence of
the mathematical proof is that the trajectory formed
from the time series is diffeomorphically related to the
actual phase-space trajectory of the dynamical system. In
order to test the mathematical properties embodied in the
diffeomorfism, that is, continuity, differentiability, inverse
differentiability and injectivity, Pecora {\it et al.} \cite{PC95} have 
developed a set of statistics intended to test quantitatively
these properties. Their algorithms are of general use and can  
in particular be applied to test topological
properties in any pair of sets of points. 
We will take advantage of this procedure by implementing
the aforementioned algorithms  to test numerically
the continuity of the logarithmic derivative of the measure
along the trajectory. 
Basically, the method is intended to evaluate,
in terms of probability or confidence levels, whether two data
sets are related by a mapping having the continuity property:
A function $f$ is said to be
continuous at a point ${\bf x}_0$ if $\forall \epsilon > 0, \exists
\delta > 0$ such that $\parallel {\bf x} - {\bf x}_0 \parallel <
\delta \Rightarrow \parallel f({\bf x})) - f({\bf x}_0) \parallel
< \epsilon$. The results are tested against the null--hypothesis,
specifically, the case in which no functional relation
between points along the trajectory and the measure exists.
This is done by means of the statistics proposed by Pecora {\it et al}
\cite{PC95}
\begin{eqnarray}
\Theta_{C^0} (\epsilon) = \frac{1}{n_p} \sum_{j=1}^{n_p}
\Theta_{C^0} (\epsilon , j)
\label{e7}
\end{eqnarray}
and
\begin{eqnarray}
\Theta_{C^0} (\epsilon, j) = 1 - \frac{p_j}{p_{max}}
\label{e8}
\end{eqnarray}
\noindent
where $p_j$ is the probability that all of the points in the
$\delta$-set, around the point ${\bf x}_j \in {\bf x}(t)$,
fall in the
$\epsilon$-set around $\frac{d{\ln \mu({\bf x}_j)}}{dt}$.
The likelihood that this will happen must
be relative to the most likely event under the null hypothesis, $p_{max}$
(see reference \cite{PC95}). When $\Theta_{C^0} (\epsilon, j)
\approx 1$ we can confidently reject the null hypothesis, and
assume that there exists a continuous function.
As in the work of Pecora {\it et al} \cite{PC95} the $\epsilon$
scale is relative to the standard deviation of the density time series,
and thus, $\epsilon \in [0,1]$.
Plots of $\Theta_{C^0}(\epsilon)$ versus $\epsilon$ can be used to
quantify the degree of statistical continuity of a given function.
In order to characterize the continuity statistics by means of
a single parameter we have also calculated,
\begin{equation}
\theta = \int_0^1 \Theta_{C^0} (\epsilon) d\epsilon
\end{equation}
\noindent The limiting values of $\theta$, namely, 0 and 1, correspond
to a strongly discontinuous and a fully continuous function, respectively.

\subsection{Summary}

A naive test to quantify noise in signals is to 
check how smooth they are. As long as more noise contaminates the signal,
more discontinuous it becomes. This is the case for example 
of additive noise, {\it e.g.} noise added to the signal. However, this
is by no means a general rule. For instance, 
intrinsic noise, that is, noise added in the equation of
motion, is not expected to affect the smoothness of the signal. 
This can be clearly seen in the case of
surrogate time series: two time series (the original series and its 
surrogate) having the same correlation structure, may have very different 
underlying dynamics, {\it i.e.} one deterministic and the other
stochastic. We can overcome the above drawback by using the trajectory
of the system, instead of a single variable.
Smoothness or continuity of the trajectory in phase space has
been used before \cite{KD92} in this context. What we propose here
is to use  the distribution of points on the trajectory (or the natural 
measure) as a way to evaluate the  the degree of 
noise in the system. Our proposal is based upon Eq. (6). This equation 
tells us that the amount of noise present in the system is directly related to
the differentiability of the natural measure, evaluated along a
typical trajectory. More noisy is the system, less
differentiable it becomes. It remains the question of how to
apply a fundamental analysis concept, like differentiability, to a magnitude
defined numerically. The "topological statistics" tools
developed by Pecora {\it et al} \cite{PC95} came to our help.
It is devised to evaluate in a statistical fashion topological
properties of functions. Eq. (8) is the statistics used  to test, against
the null hypothesis, the degree of continuity of our function. As long
as this statistic approaches  unity we are confident that we have
a continuous function. We must note
that we have evaluated the continuity of the numerical derivative
of the measure, which is
equivalent to test differentiability, a computationally easier procedure.
Continuity is directly related to the resolution with which
we are looking at the function, that is $\epsilon$, so the
statistics is actually $\epsilon$--dependent $\Theta_{C^0} (\epsilon)$. In
order to provide us with an overall continuity test, we summed up
over the whole range of $\epsilon$ obtaining a single parameter
$\theta$ (see Eq. (9)) which we use hereafter to characterize continuity.  

\section {Models and Numerical Procedures}

Here we discuss the various stochastic dynamical systems on which we have 
investigated the efficiency of our approach. The numerical procedures followed
to reconstruct the space from time series related to some of the
coordinates of those models and to evaluate the statistical differentiability
of the natural measure are also described. We also discuss a variant of
the surrogation process which consists of producing partially 
surrogate series. The method is a way to vary the degree of stochasticity
of the time series.

\subsection {The stochastic Van der Pol Oscillator}

The simplest case in which we can apply our ideas is in
a nonlinear oscillator. We have used the Van
der Pol (VdP) oscillator \cite{VV28} with an additive stochastic term.

\begin{eqnarray}
{\dot x} & = &  y + \eta G_1 \nonumber \\
{\dot y} & = &  -(x^2 - 1) y - x +  \eta G_2 \\
\label{e9}
\nonumber
\end{eqnarray}

\noindent The parameter $\eta$ represents the noise
level, and $G_i(t)$ are uncorrelated Gaussian noises, such
that $G_i(t) \in$ Normal(0,$\sigma$), zero mean and
standard deviation $\sigma$.
Without loss of generality we take $\eta=1$,
and tune the degree of noise only by the standard deviation $\sigma$.
 
Numerical integration was carried out by means of an Euler 
algorithm. 20000 data points have been generated. 
Using the $x$-coordinate as our ``experimental" time
series, we have made a reconstruction with an embedding
dimension in the range 2--20, and a $\tau$ lag of 
10 (in sampling units). Every reconstruction has been
rescaled to the unit hyper-square.

Density time series have been obtained for each 
reconstruction. Starting with the first point in the
reconstructed phase space, we follow the trajectory 
recording the density of points around each of the
trajectory's point. In order to estimate this 
density, we have used the Epanechnikov kernel \cite{Ep69}, which,
roughly speaking,``weighs" the points according to its
distance to the center. This is preferable to the
Gaussian kernel, which is of infinite support (and
therefore has a lower computational efficiency) and the
``square" kernel which gives equal status to the different
points in the ball around the reference point. 
A parameter that has to be chosen carefully is the ball radius
used to estimate the density. If the radius is
too small, the low density regions will
be practically depopulated and the measure will be underestimated. 
On the other hand, if the radius is too big, the estimation will 
capture points which are not really part of
the neighborhood of the reference point. Of course,
the ball radius is a function of the data 
points being used in the reconstruction process and
must be chosen according to this fact. We have
used a radius of the ball of 5\% of the total
attractor extent.
In evaluating the continuity statistics, we average $\Theta_{C^0}
(\epsilon , j)$ over $n_p$ points (see Eq. (6))
randomly distributed in the trajectory, typically $10 \%$ of the
total record.

Fig. \ref{fig_3} shows a typical density time series from the
$x$-coordinate of the Van der Pol oscillator.
Albeit qualitative, the smooth behavior 
of this density along the trajectory is readily noted.

\subsection{The stochastic Lorenz System}

The VdP oscillator will allow us to get a closer look at the 
procedure we are implementing. However, a nonlinear oscillator
is a somewhat trivial example and we want to apply the
method to more complicated cases. In fact, the ultimate 
objective of the methodology is to discriminate random
behaviour from a deterministic one, and this is
specially important in the case of chaotic behaviour.

The Lorenz system \cite {Lo63} with an additive stochastic term is an
adequate choice. The related system of differential equations  can be
written as:
\begin{eqnarray}
{\dot x} & = & -s x + s y + \eta G_1 \nonumber \\
{\dot y} & = &  -y + rx - xz + \eta G_2 \\
\label{e10}
\nonumber
{\dot z} & = & -bz + xy + \eta G_3 \nonumber
\end{eqnarray}
The parameters used in the calculations
are, $s$ = 10.0, $r$ = 28.0 y $b$ = 2.66, which give chaotic behaviour
in the case $\eta = 0$. $G_i(t)$ and $\eta$ were defined above.
When not specified the results for the Lorenz system discussed in the following 
Section were obtained for an embedding dimension of 3, which is greater than
the correlation dimension of the Lorenz system. 

Numerical integration of the Lorenz system was carried out by means
of the Euler method. The time integration step was 0.01.
Time series with 16384 data points and
their respective surrogates were subsequently generated.
The reconstruction was performed by the usual time--delay method
\cite{Ta81,PC80,SY91},
with a time delay given by the first zero of the autocorrelation
estimate (10 in units of the integration step)
on an embedded phase space of dimension 3.
The natural measure $\mu({\bf x}(t))$
along the trajectory was calculated by means of the Epanechnikov kernel
density estimator \cite{Ep69} with an sphere of radius 5\% of the
attractor extent. As in the VdP oscillator the continuity statistics
was evaluated including up to 10\% of the points in a given record.

The sensitivity of the time derivative of the measure to stochasticity
is illustrated in Fig. \ref{fig_4}.
This figure shows that whereas the surrogate
of the $x$--coordinate time series is as "smooth" as the original series,
the time derivative of the logarithm of the measure is much more spiked 
in the surrogate
than in the original series.

\subsection{The Mackey--Glass model}

In order to investigate the effects of the embedding dimension
we have also considered the high dimensional system introduced in
\cite{MG77}. The dynamical system is described by means of
the following delay--differential equation,
\begin{equation}
{\dot x} = \frac{ax(t-\delta}{1+x(t-\delta)^c}-bx(t)
\end{equation}
\noindent This equation has been proposed to model nonlinear feedback
control in physiology.
We use the set of parameters that gives an attractor
dimension of $\approx 7.5$ \cite{KD92}, namely, $a=0.2$, $b=0.1$, $c=10$,
and $\delta = 100$. The resulting time series were
analyzed with a time
delay given by the first zero of the autocorrelation estimate,
and the measure was evaluated on spheres of radius 10\% of the attractor
extent and the continuity statistics with 10\% of the points in the
total record. 

\subsection{Partial surrogation}

The surrogation process is a well established method in the context
of nonlinear time series analysis \cite{TE92,Sc98}. 
In typical applications a single time series is available. From this time
series, an ensemble of the so-called surrogate series are generated that mimic
certain properties of the original. For example, by simply 
scrambling the temporal order of the points in the original, one
obtain surrogate time series which preserve the mean, variance, etc.
One of the most popular methods of producing surrogate time series
consists of  shuffling the phases in the Fourier transform
of the original data set \cite{TE92}.
In this way, each value of the Fourier transform
of the original data is multiplied by a random phase $\exp(i{\phi})$,
with $\phi \in$ random $[0,2\pi]$ \footnote{In order to
get a real time series in the antitransformation we multiply 
symmetrically with respect to the center of the transform}. 
The procedure generates  a new time series with
the autocorrelation structure of the original. 
Here we propose to introduce
an additional factor, $\exp(i{\phi}{\alpha})$, with $\alpha \in [0,1]$.
This factor allows us to control the degree of stochasticity by tuning
the parameter $\alpha$.

\section {Results}

\subsection{Sensitivity of the time derivative of the measure to stochasticity}

In order to test the efficiency of the approach proposed here we have
first evaluated the continuity statistics either on a coordinate or on
the time derivative of the natural measure time series of the stochastic Van der Pol
oscillator.  Fig. \ref{fig_5} shows the continuity
statistics for the $x$--coordinate and for the time derivative of 
the reconstructed natural measure 
with and without noise. There
is almost no modification in the statistics of the  
coordinate upon  noise addition. However,  
the statistics of the density time series reflects
very clearly the presence of the stochastic term.
This illustrates the efficiency and novelty of our approach and supports its
application to more complex cases. We must remark that we are using
the commonly named "dynamical" noise, that is, a stochastic term
added in the dynamical equations, instead of the "measurement" 
noise, which is added after the "clean" integration step.
Preliminary results shows that an efficient method to discriminate
both types of noise, can be achieved by using the continuity
statistics over the density and the coordinates, but this 
deserve further research, and eventually will be published
elsewhere.

The results for the continuity statistics
of the time derivative of the reconstructed measure from the
$x$--coordinate of the Lorenz system are illustrated in Fig. \ref{fig_6}.
The results of Fig. \ref{fig_6}a show that the time derivative
of the measure
in the original series is "more continuous" (in a statistical sense) than
its surrogate. Partial surrogation ($10\%$)
decreases the degree of continuity of the time derivative of the measure
in an extent lower than total surrogation, as expected.
On the other hand, the results show that the continuity of the
totally surrogated series shows almost no dependence on whether it
has been derived from the original series or from a partially surrogated
series (10 \% surrogated).

The results for the stochastic Lorenz system reported in Fig. \ref{fig_6}b
clearly show that the stochastic terms  significantly decrease the 
statistical continuity of
the time derivative of the measure. Surrogation of the stochastic
series produces a further decrease of continuity, indicating that the
series still has some degree of determinism. The degree of stochasticity
of a time series can be quantified by
calculating the integral of the continuity statistics as defined in Eq. (8).
Fig. \ref{fig_7} a) shows how steeply
$\theta$ decreases with the percentage of surrogation. Similarly, $\theta$
decreases with the standard deviation of
the Gaussian noise in the stochastic Lorenz system (Fig. \ref{fig_7}b), as expected.
Thus, the magnitude $\theta$ can be used to evaluate the relative
stochasticities of a set of experimental time series.

\subsection{Dependence on the Embedding Dimension}

A point of crucial relevance is how the above results change with the
embedding dimension $m$. We have investigated this question on the
Lorenz system and on the high--dimensional system discussed in III.D 
\cite{MG77}.
The results for the Lorenz system depicted in Fig. \ref{fig_8}a show
that $\theta$ decreases with the embedding dimension. This is a
consequence of working with a fixed sphere radius for all $m$ and
of the numerical noise that should increase with $m$. The decrease of
$\theta$ is stronger in the surrogate series, although it is
likely that the difference between the two should decrease for large
enough $m$. In any case, the difference in $\theta$ between the original
and the surrogate series
changes only from 0.35 to 0.49 when $m$ is varied in the range  3--10.
The behavior of $\theta$ in the high--dimensional system is far more
intricate (see Fig. \ref{fig_8}b). For $m$ well below the attractor dimension
the measure for the
surrogate series seems to be more continuous than that for the original series.
The reason for this rather odd behavior has to be found in the heavy crossing
of trajectories that occur at $m$ far below the attractor dimension
\cite{AB93}.
In those cases, surrogation seems to have a smoothing effect.
Instead, for $m > 7$ the behavior is similar to that of the Lorenz
system, although the difference between the original and the surrogated
records is substantially smaller. This point further study that
is actually in progress.

Fig. \ref{fig_9} illustrates the dependence of the continuity 
statistics of the measure
on the embedding dimension $m$ for the stochastic Van der Pol oscillator.
In this case the simplicity of the attractor gives an almost null
dependence on $m$ for no noise or very low noise levels. When
the noise is increased the behavior is in line with that found
in the Lorenz system, namely, a decrease of $\theta$ as $m$ increases.
This dependence on $m$ is more noticeable the greater 
the noise level. 

The study of the dependence of the continuity statistics of the measure derivative
on the embedding dimension involves numerical difficulties that deserve some comments. 
In the case of low enough
embedding dimension and moderate number of data points, the estimation
can be achieved confidently for almost all the points in the
trajectory. Of course, "moderate" and "long enough" are terms which depend on
the attractor dimension. We think that the same criteria 
followed to get for example a reliable estimation of the 
correlation dimension \cite{AB93}, apply in this case.
In our approach we have introduced noise explicitly, adding
another factor to be considered in the estimation step.
As it is well known, as "extra" dimensions are included in
the embedding process, noise populates them more or less
uniformly. This is specially problematic in the case
of "simple" systems, like the VdP oscillator, as the trajectory in a 
stochastic oscillator
may "wander" far from the zero-noise limit cycle. In these
excursions, the measure swept by the trajectory is unavoidably
constant, because no other points are in the neiborhood of the
evolution, except those which are time correlated. In such a case,
a constant measure results, and thus, a high value of the
continuity statistics. This effect is more noticeable as the embedding
dimension is increased, because "more" space is available 
(see Fig. \ref{fig_10}).

\subsection{Application to Mixed Time Series}

A distinctive feature of our method is the possibility of
using it in different ranges of a given time series. In this
way we can examine short records and evaluate its stochasticity.
Bearing this in mind we have devised the following
example:
Suppose we have a time series which is half deterministic and
half stochastic. Could our method discriminate both
behaviors in the same time series?.
In order to answer this question, we have generated a
single time series (16384 points) with the first
half coming from the $x$-coordinate of the deterministic Lorenz
system, and the second half coming from its surrogate
(100 \% randomization) time series. We have applied the continuity
statistics over four regions in the time derivative of the density record
(two randomly selected in the first half and two in the second half).
Fig. \ref{fig_11} shows the results. It is then clear that the
statistics utilized here can discriminate stochastic from
deterministic behaviour.
Fig. \ref{fig_11} also shows
the statistic for the whole time series (same number of
reference points randomly selected along the time series).
The results are midway between those for the stochastic and deterministic
ranges. 

As another example of the use of our methodology we have 
investigated the effect of a noise burst in the Van der Pol oscillator. The
dynamic equations are rewritten as,

\begin{eqnarray}
{\dot x} & = &  y + A(t) \eta G_1 \nonumber \\
{\dot y} & = &  -(x^2 - 1) y - x +  A(t) \eta G_2 \\
\label{e11}
\nonumber
\end{eqnarray}

\noindent where $A(t) = 1$ if $t_1 < t < t_2$ and $0$ otherwise. The interval
$[t_1, t_2]$ is a small interval where we "turn on" the stochastic
term. The idea behind this system is to test the capabilities
of the method to detect the noise introduced. Using 20000 data
points, we have used 500 consecutive points with the stochastic
term added. In Fig. \ref{fig_12}  we show a typical time series (and
the measure time derivative) where noise has been turned on in the
time interval 10000--10500, with a strength of $\sigma=0.05$. 
Fig. \ref{fig_13}
shows Pecora et al. \cite{PC95} statistics applied to five regions
of the whole series, one of them being the stochastic region.
Again, our method clearly discriminates  noisy and clean regions.

\subsection{Application to Experimental Time Series}

In order to test our method in actual experimental time series,
we have used the data set from the Belusov-Zhabotinski (BZ) chemical 
reaction \cite{RS83}. As  shown by those authors the
apparently random behaviour of the amplitude of the concentration
of bromide ions can in fact be explained by deterministic laws. 
We will use here our approach to confirm the above finding. 

Using the bromide concentration time series, as in the work of 
Roux {\it et al.} \cite{RS83} we repeat the procedure explained
above. We have used a $\tau$ = 30 (in sampling units) for the
reconstruction and an embedding dimension of 4.
Fig. \ref{fig_14} shows the continuity statistics for both 
the time derivative of the reconstructed measure using the BZ time 
series and for its 100 \% surrogate. The large difference between the two
and the rather high value of $\Theta_{C^0}$ confirm our proposal in 
the sense  that one can explain the behaviour of this record as the 
output of a deterministic dynamical system.

\section {Concluding Remarks}
In brief, we have proposed a method to identify determinism in
time series which exploits the continuity of the logarithmic time derivative
of the natural measure along the trajectory, that is, its differentiability. 
The method is based upon a
formula which explicitly shows the sensitivity of the measure to stochasticity.
In the present work we have adapted the statistical method of
Pecora {\it et al.} \cite{PC95} to investigate the continuity of the 
time derivative of the measure.
Results of partially
surrogated series and series derived from two stochastic dynamical systems
and a high--dimensional system,
clearly illustrate the suitability of the present
method to the problem at hand. As we have shown, the method is 
very easily applied to any kind of time series, from the simplest
one, as is the case of limit cycle oscillators, to the high--dimensional
cases. Given a time series, and once done the standard reconstruction process,
obtaining the density (or its time derivative) is a straightforward
procedure. Then, one can use Pecora {\it et al.} approach over the whole
time series or in selected pieces of it. 
A warning against a blind application of the method is in order,
particularly in what concerns the evaluation of the continuity of the measure 
time derivative. A careful inspection of the density time series must 
be done before any further operation is performed.

The dependence
of the continuity statistics on the embedding dimension in low and
high--dimensional systems, indicate that applications to real (experimental)
time series would eventually require a thorough investigation
of this point in each particular case, as is common in time series analysis.
In any case, the fact that the method works reasonably well
on short time series, supports its usefulness for the analysis of
experimental series. We have shown this in a simple case as it is
the Belousov-Zhabotinski reaction, confirming previous findings.

In a broader sense, the application of continuity statistics over
the density time series is a new aspect of the possibilities
offered by the Lagrangian Measures \cite{Or96}. As we have
shown previously, the use of more traditional tools on this density,
such as Fourier Transforms or histograms, may
help in extracting new information on the underlying dynamical
system. 

\acknowledgments
Thanks are due to E. Hern\'andez, M. SanMiguel and R. Toral for many
useful comments and suggestions.
This work was supported by grants of the spanish CICYT (grant no. PB96--0085),
the European TMR Network-Fractals c.n. FMRXCT980183, and 
of the Universidad Nacional de Quilmes. G. Ortega is member of 
CONICET Argentina and also thanks the "Generalitat Valenciana"
for support.

\newpage

\begin{figure}[tbh]
\caption {Illustrates two alternative evolutions in dissipative 
systems: Liouvillian evolution (upper), and, Lagrangian
evolution (lower).}
\label{fig_1}
\end{figure}

\begin{figure}[tbh]
\caption {Scheme followed to analyze a dynamical system and to
calculate the natural measure $\mu_{\pi}$ on the reconstructed space.}
\label{fig_2}
\end{figure}

\begin{figure}[tbh]
\caption {Typical records of the $x$--coordinate (continuous line) and of
the natural measure (broken line) time series (in arbitrary units) from
the $x$--coordinate of the Van der Pol oscillator. The results correspond
to the parameters given in
the text whithout stochastic term, and an embedding dimension of 4}
\label{fig_3}
\end{figure}

\begin{figure}[tbh]
\caption {Illustrates the differences between using a coordinate or the
time derivative of the natural measure in the evaluation of the continuity 
statistics. The results correspond to the Lorenz system without noise.
a) $x$--coordinate (broken line) and its $100\%$ surrogate
(continuous line). b) derivative of the reconstructed natural measure 
(broken line) from the $x$--coordinate and from the $x$--surrogate
surrogate (continuous line).}
\label{fig_4}
\end{figure}

\begin{figure}[tbh]
\caption {Van der Pol oscillator: continuity statistics (see Eq. (\ref{e7}))
for the $x$--coordinate (lower frame) and for the time derivative of the
measure along the trajectory obtained from the $x$--coordinate (upper frame).
Results for the deterministic system (broken
curves) and the stochastic system with $\sigma=0.06$ (continuous curves)
are shown.}
\label{fig_5}
\end{figure}

\begin{figure}[tbh]
\caption {Continuity statistics (see Eq. (\ref{e7})) for the time
derivative of the measure along the reconstructed trajectory from
the $x$--coordinate
of the Lorenz system. a) Results for the original time series, and
for the series partially ($10\%$) or totally surrogated ($100\%$), and
a combination of both. b) results for the Lorenz system with noise (see
Eq. (8)) and for its surrogate series.}
\label{fig_6}
\end{figure}

\begin{figure}[tbh]
\caption {Integral of the continuity statistics
(as defined in Eq. (8)) for time series derived from the Lorenz system.
The results correspond to: a) partially randomized series with increasing degree
(percentage) of randomization, and, b) stochastic Lorenz system with increasing
standard deviation of the Gaussian noise in Eq. (8).}
\label{fig_7}
\end{figure}

\begin{figure}[tbh]
\caption {Integral of the continuity statistics (see Eq. (8)) as
a function of the embedding dimension for time series (filled symbols),
and their surrogates (empty symbols), derived from,  a) the Lorenz system,
and, b) the high--dimensional system proposed in Ref. {\protect
\cite{MG77}}. The error bars in the results for the surrogate series account
for averages over 5 realizations.}
\label{fig_8}
\end{figure}

\begin{figure}[tbh]
\caption {Integral of the continuity statistics (see Eq. (8)) as
a function of the embedding dimension for time series
derived from the the $x$--coordinate of the Van der Pol oscillator 
with various degrees of noise:
$\sigma$ = 0 (filled circles), 0.01 (open circles), 0.03 (filled squares),
0.06 (open squares) and 0.09 (filled triangles).}
\label{fig_9}
\end{figure}

\begin{figure}[tbh]
\caption {Measure estimate in the case of a stochastic
Van der Pol oscilator
($x$-coordinate) for a embedding dimension of 19, and
$\sigma$ = .5}
\label{fig_10}
\end{figure}

\begin{figure}[tbh]
\caption {Continuity statistics for the time derivative of the measure
corresponding to the $x$--coordinate of
the Lorenz system. The time series was formed by
joining the original time series (first half) to its fully
randomized series (second half). See text.}
\label{fig_11}
\end{figure}

\begin{figure}[tbh]
\caption {$x$--coordinate (broken line) and derivative of the natural
measure (continuous line) time series of the Van der Pol oscillator in
which a noisy burst, of $\sigma=0.05$ has been introduced in the
time interval 10000-10500.}
\label{fig_12}
\end{figure}

\begin{figure}[tbh]
\caption {Continuity statistics for the time series of Fig. \ref{fig_12},
applied to five regions of the series one of which is the stochastic region
(broken line). The noise intensity is $\sigma=0.05$}
\label{fig_13}
\end{figure}

\begin{figure}[tbh]
\caption {Continuity statistics for the
BZ reaction time series (solid line) and its surrogate
(broken line)}
\label{fig_14}
\end{figure}

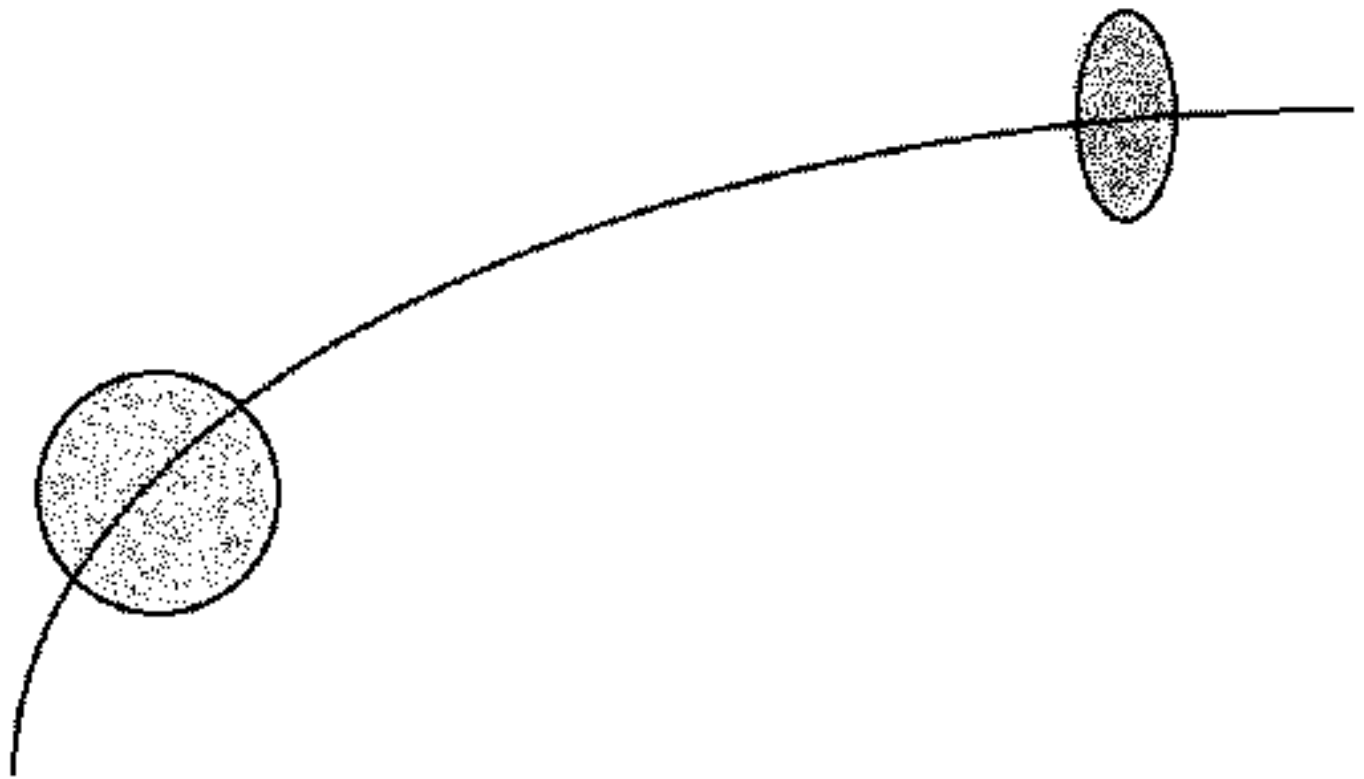
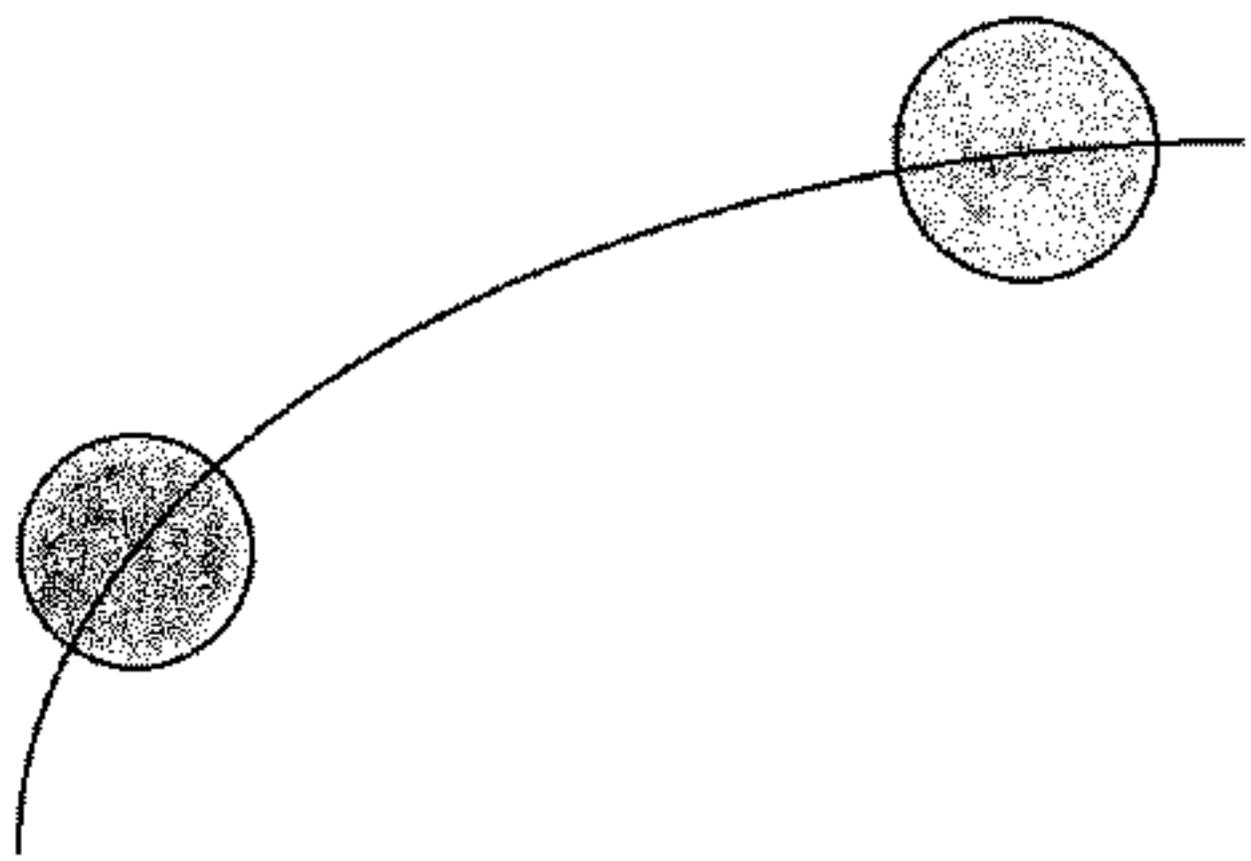
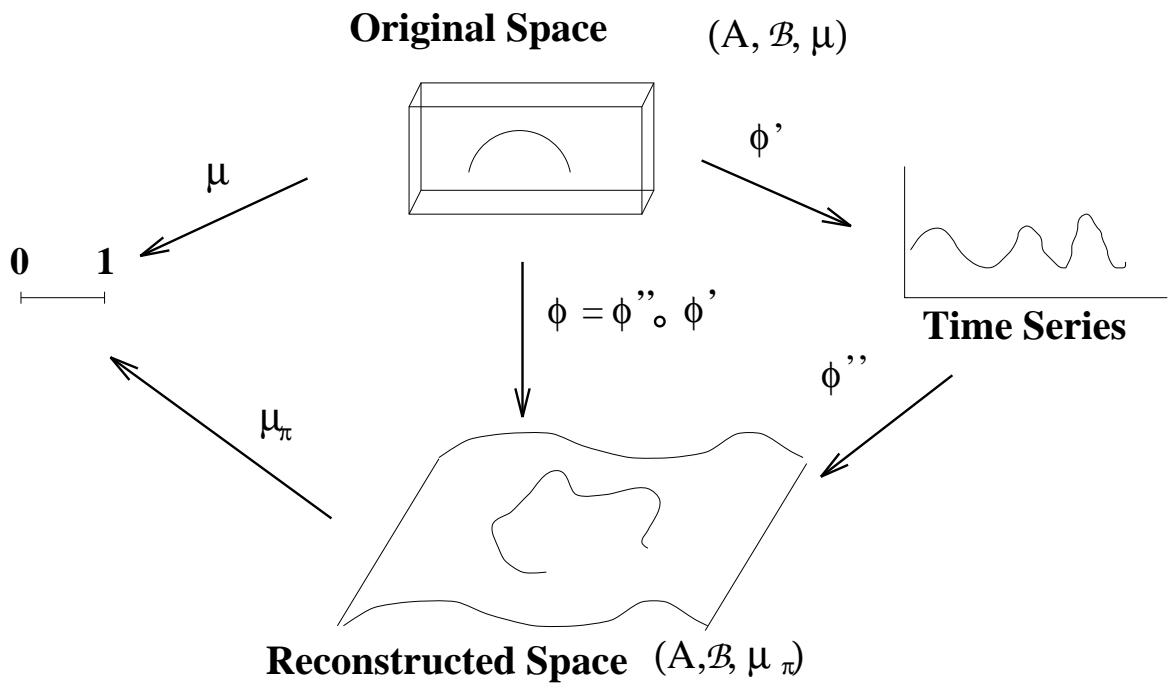
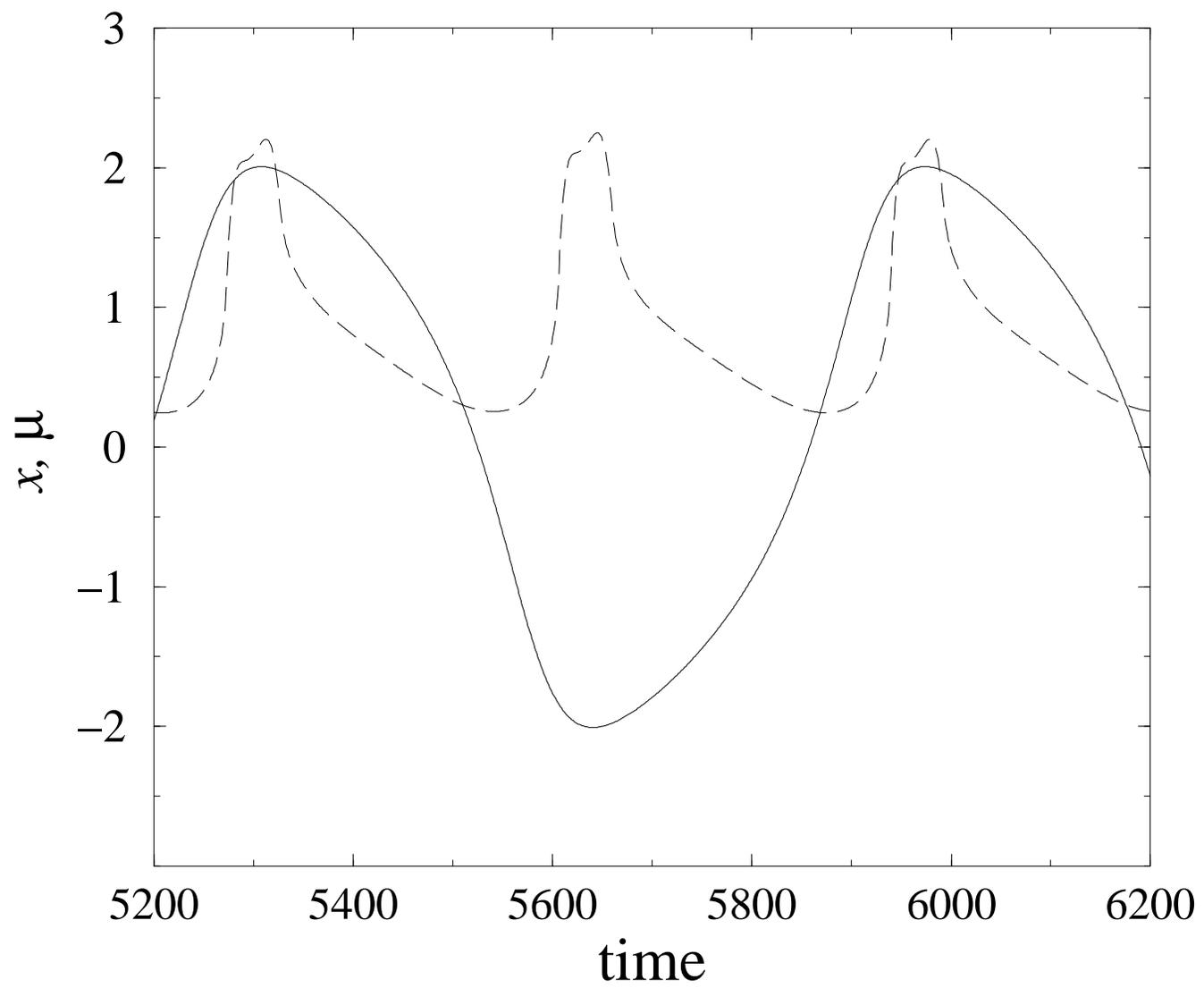
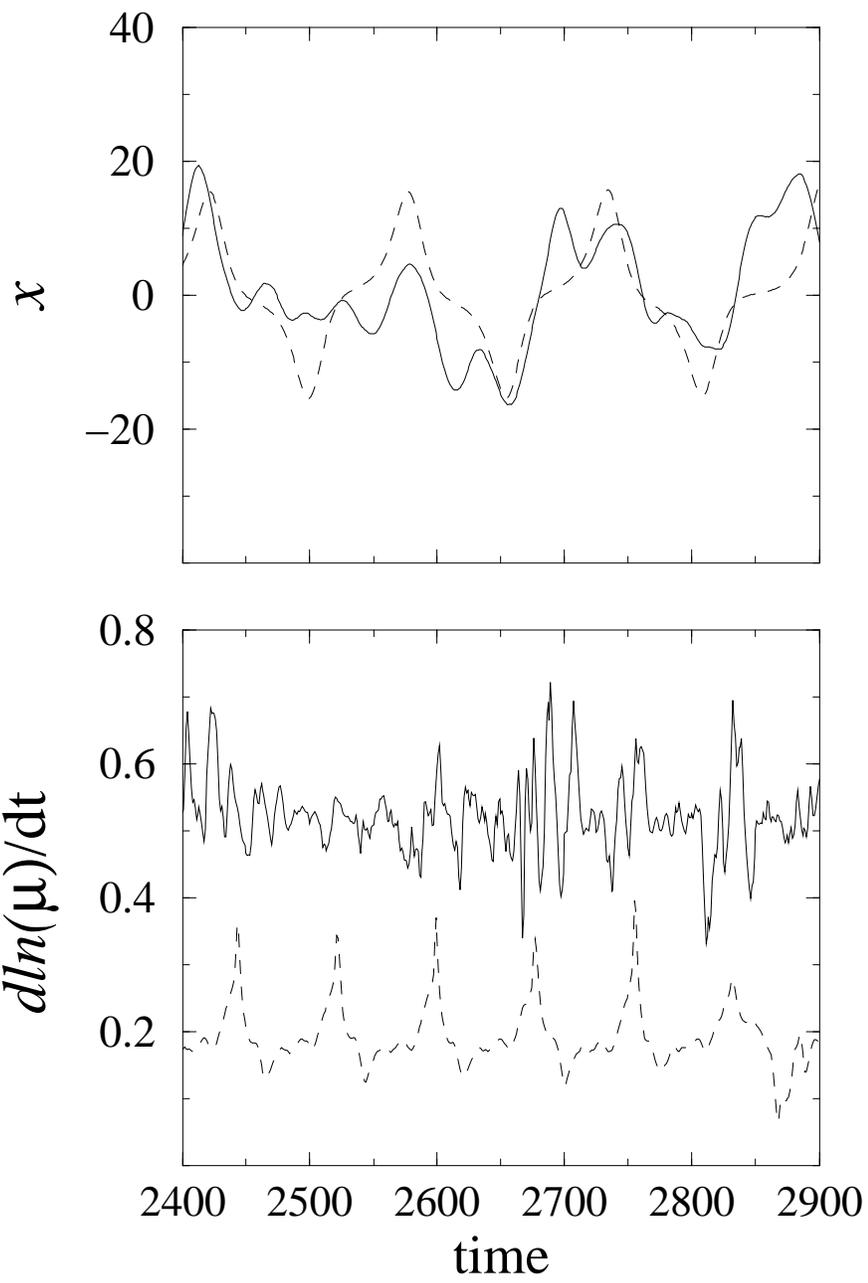
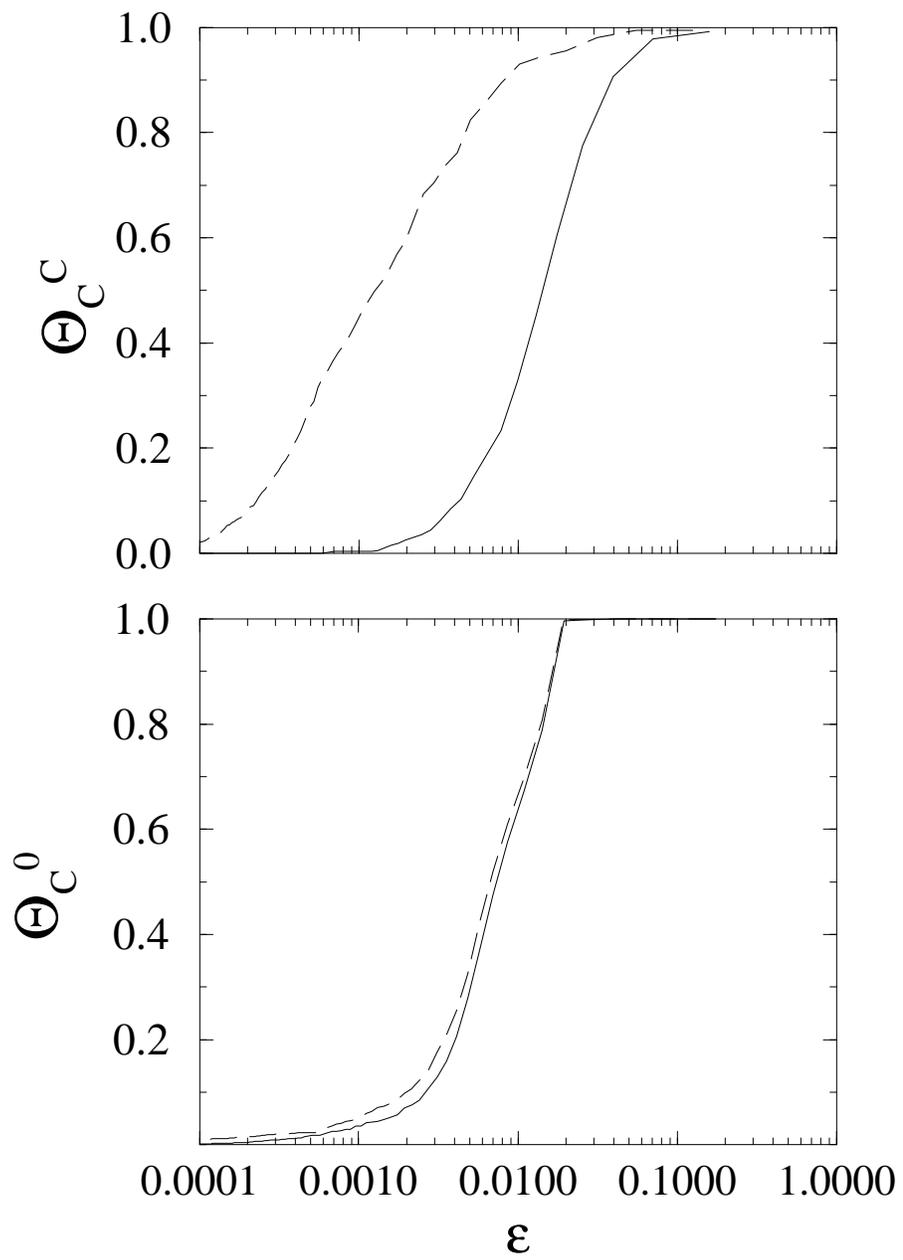
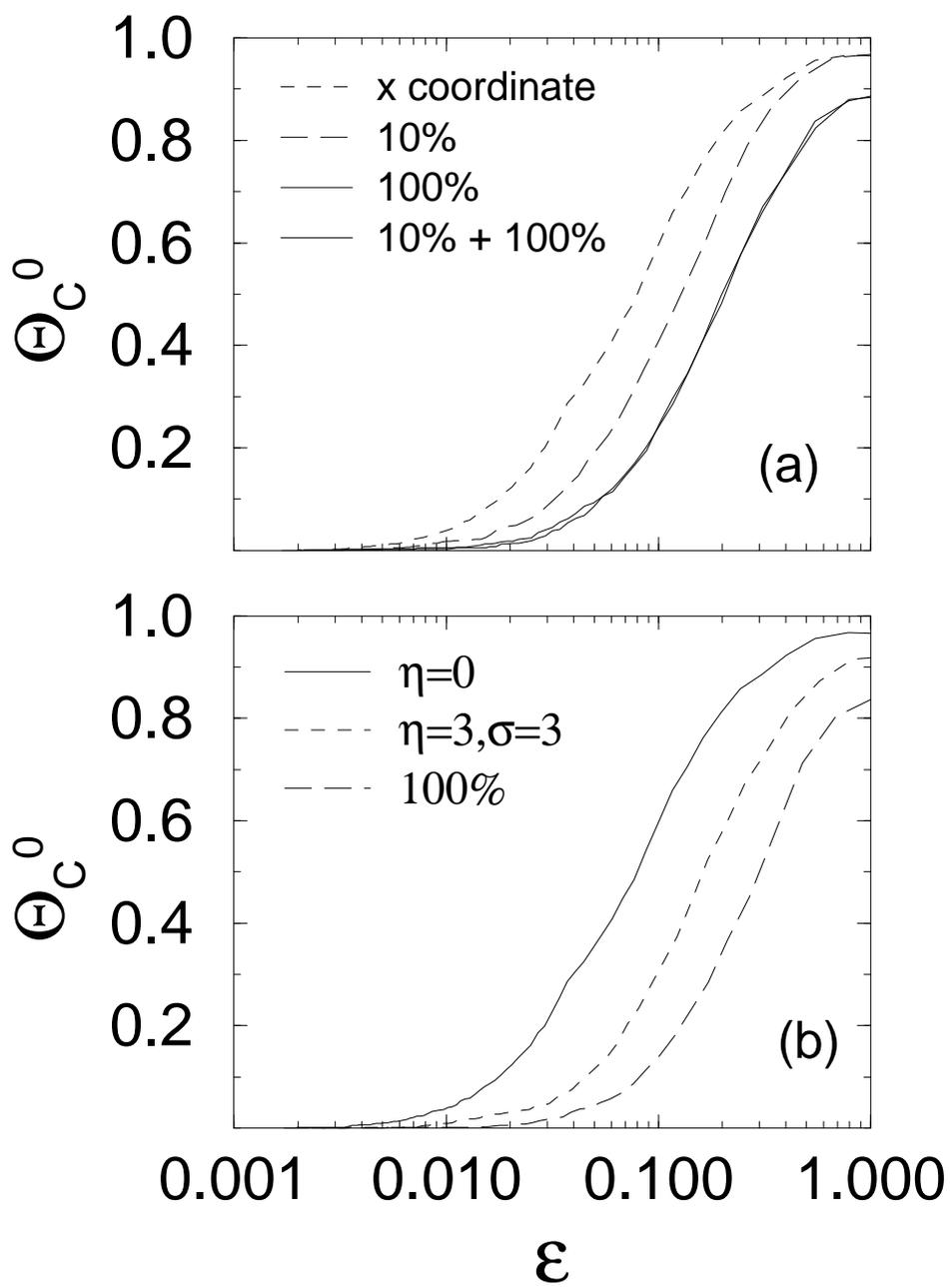
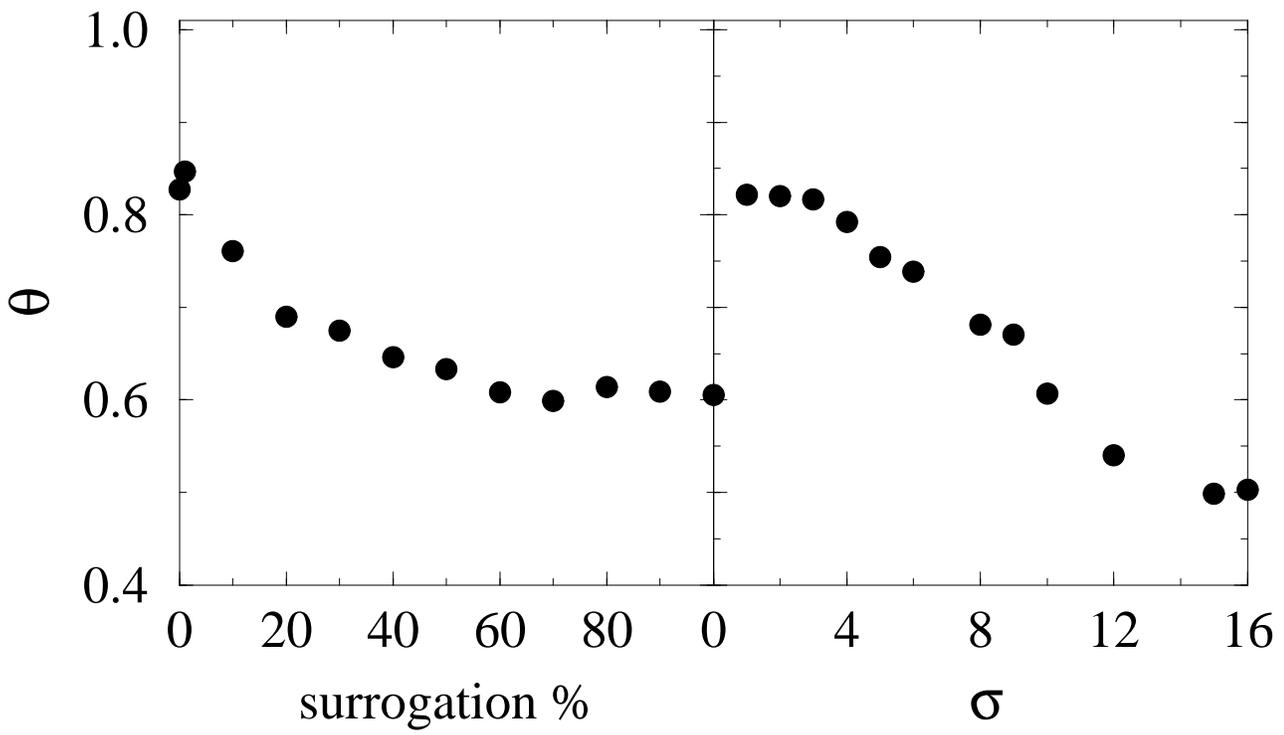
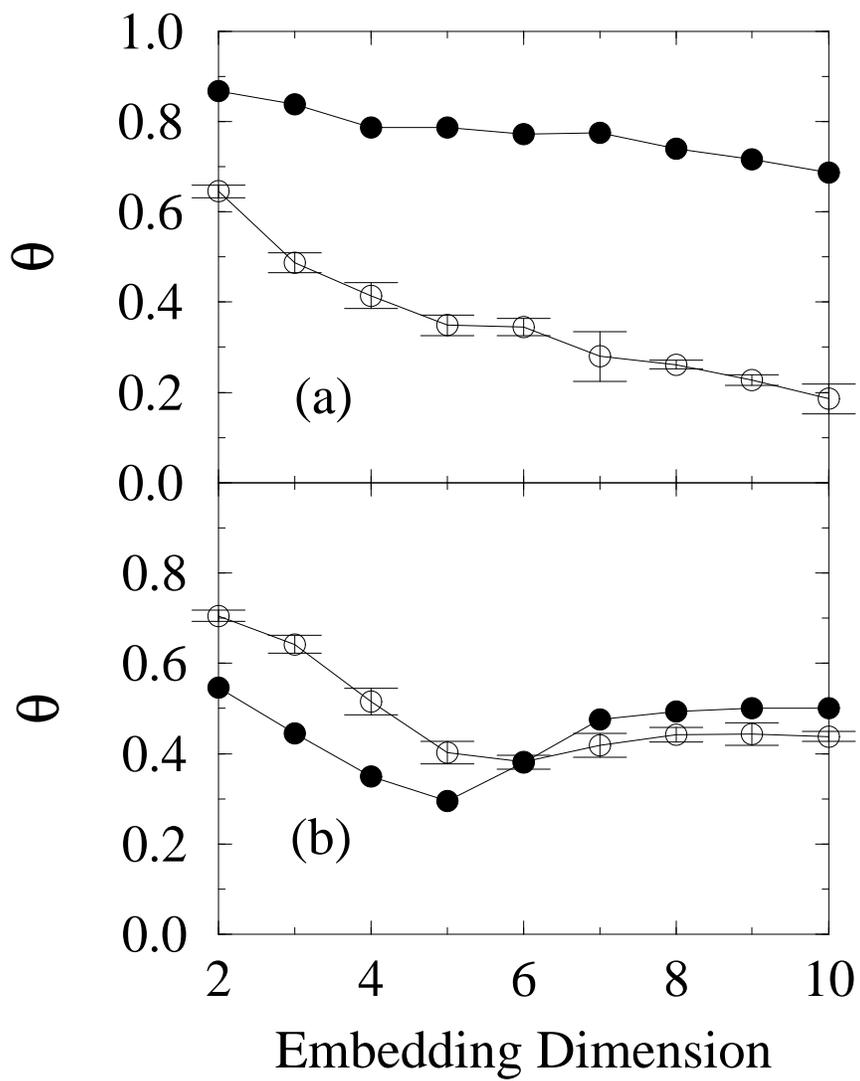
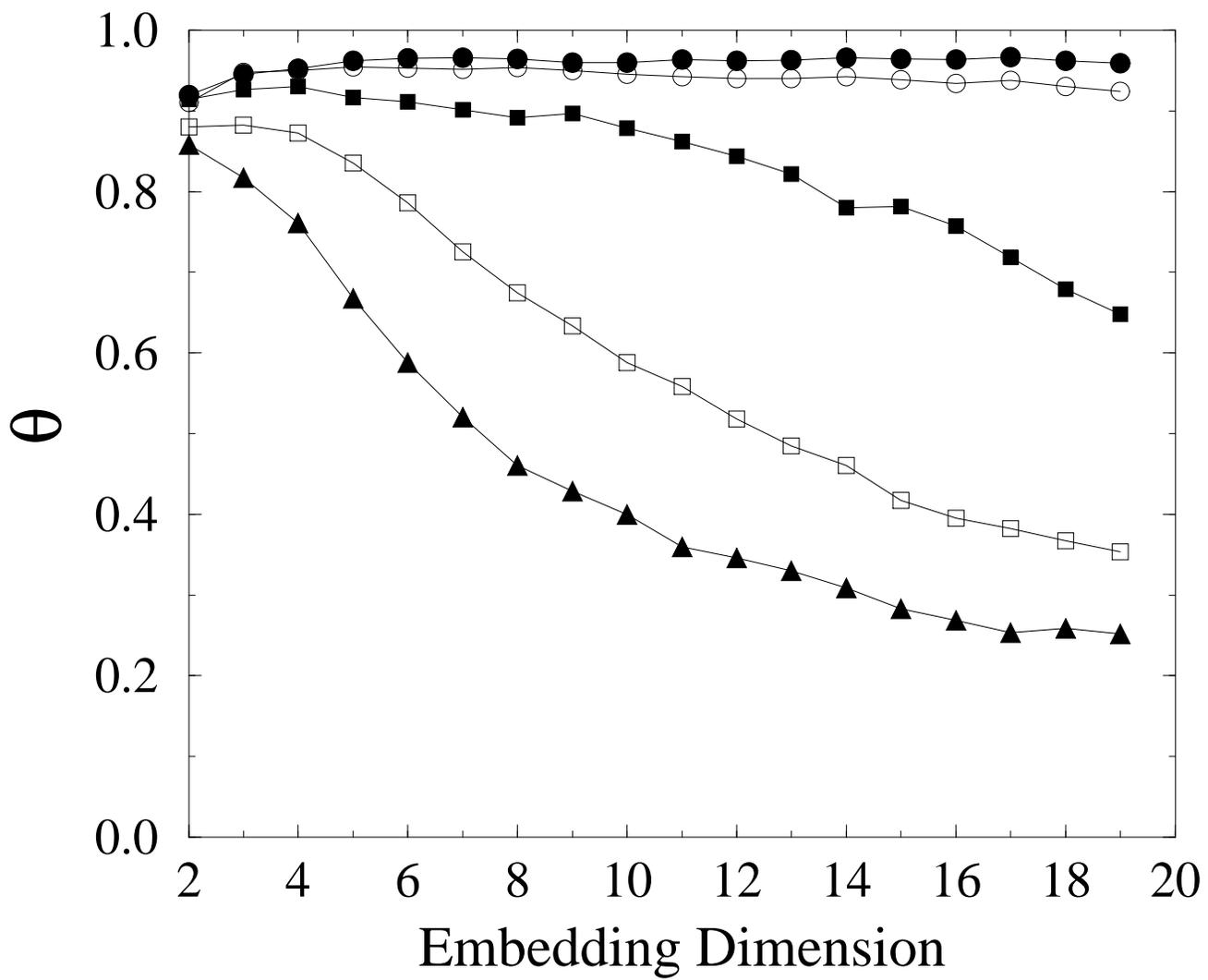
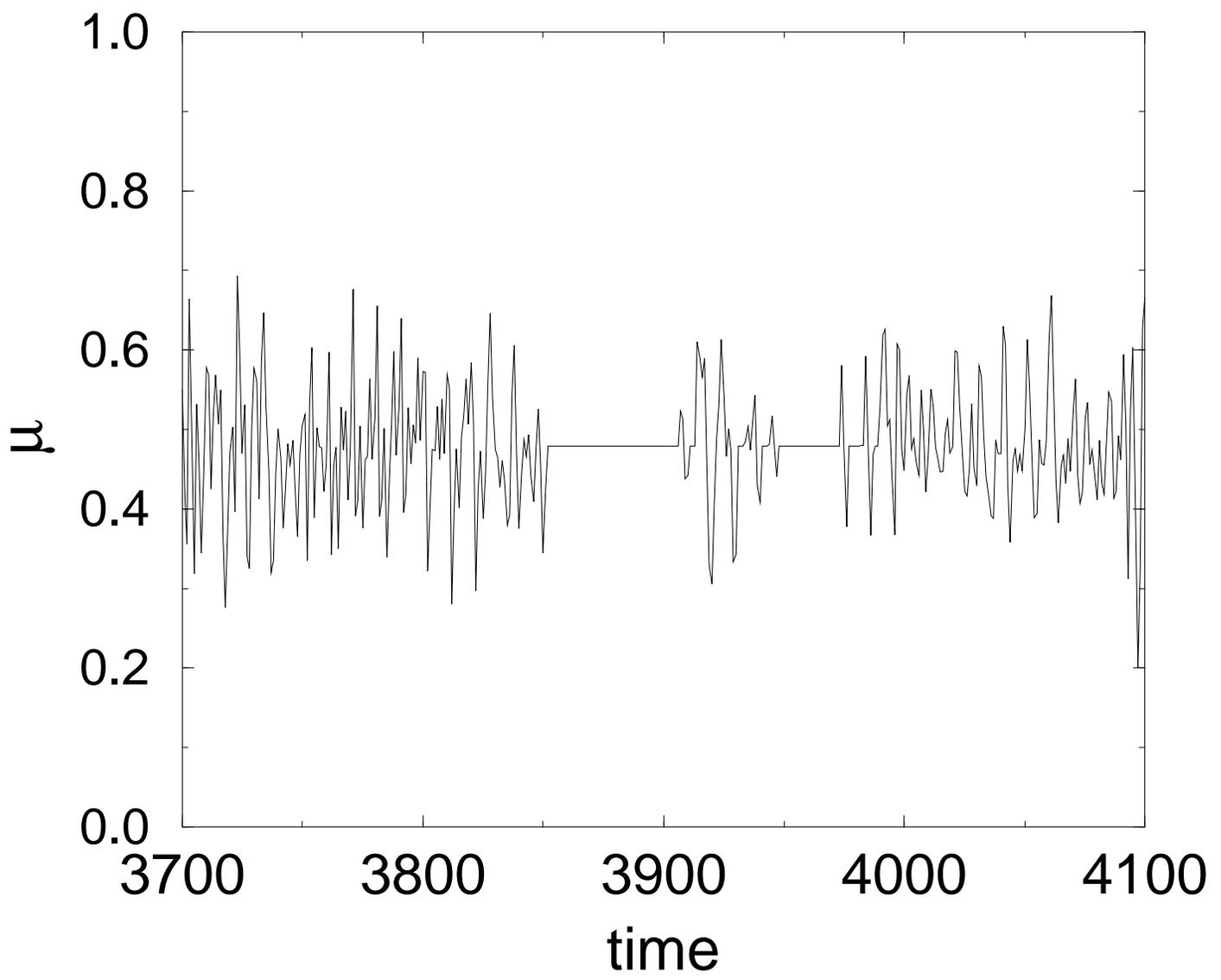
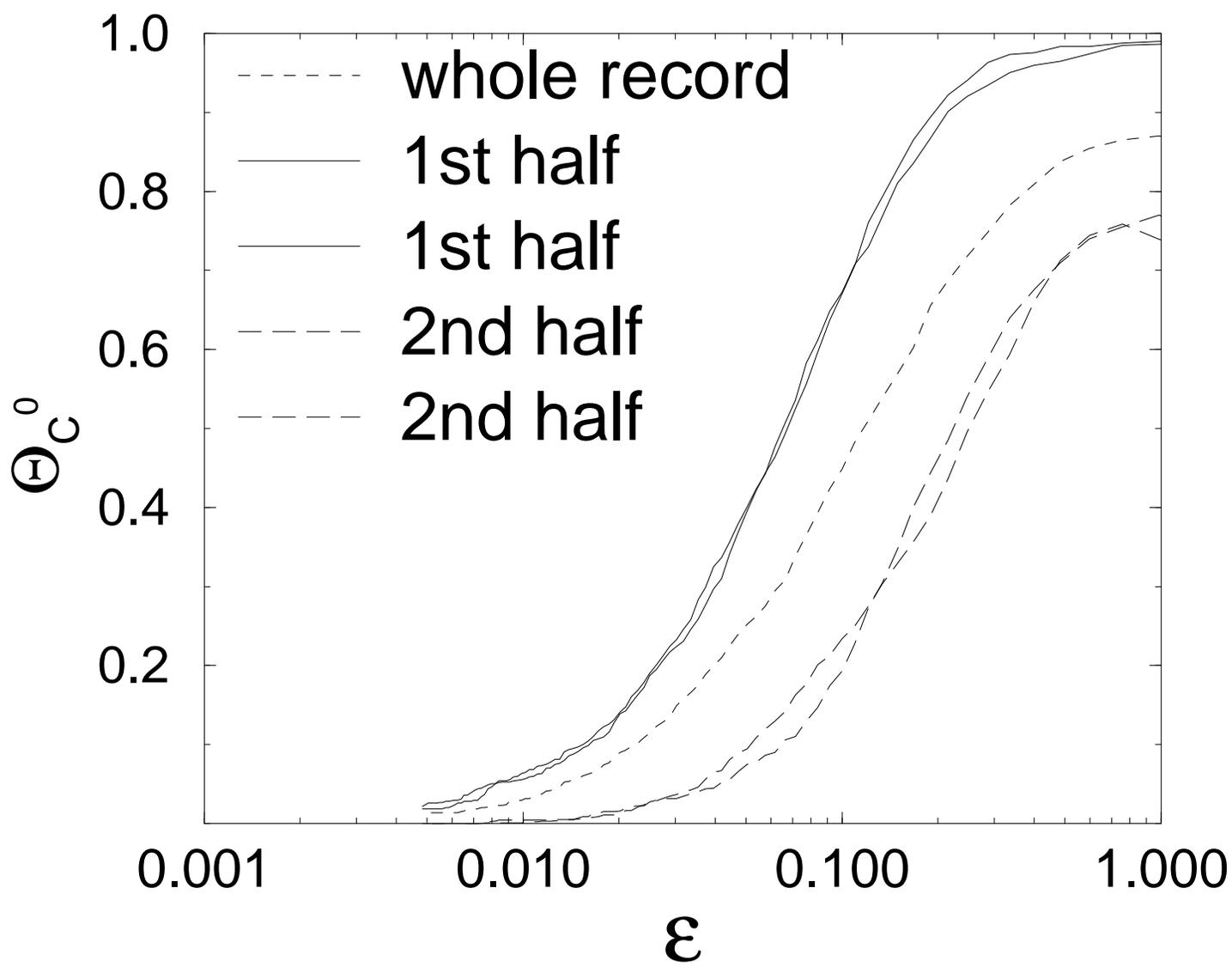
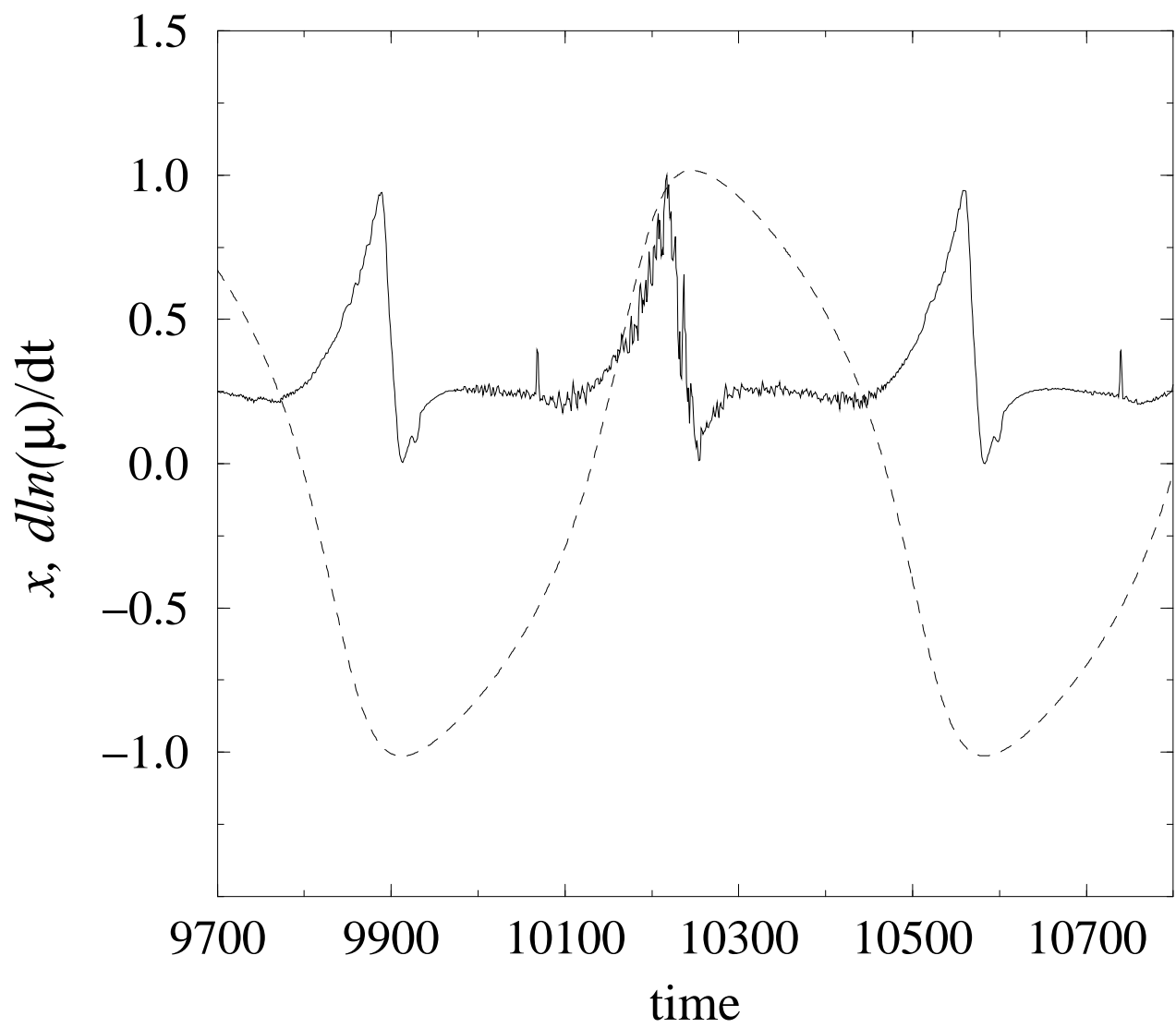
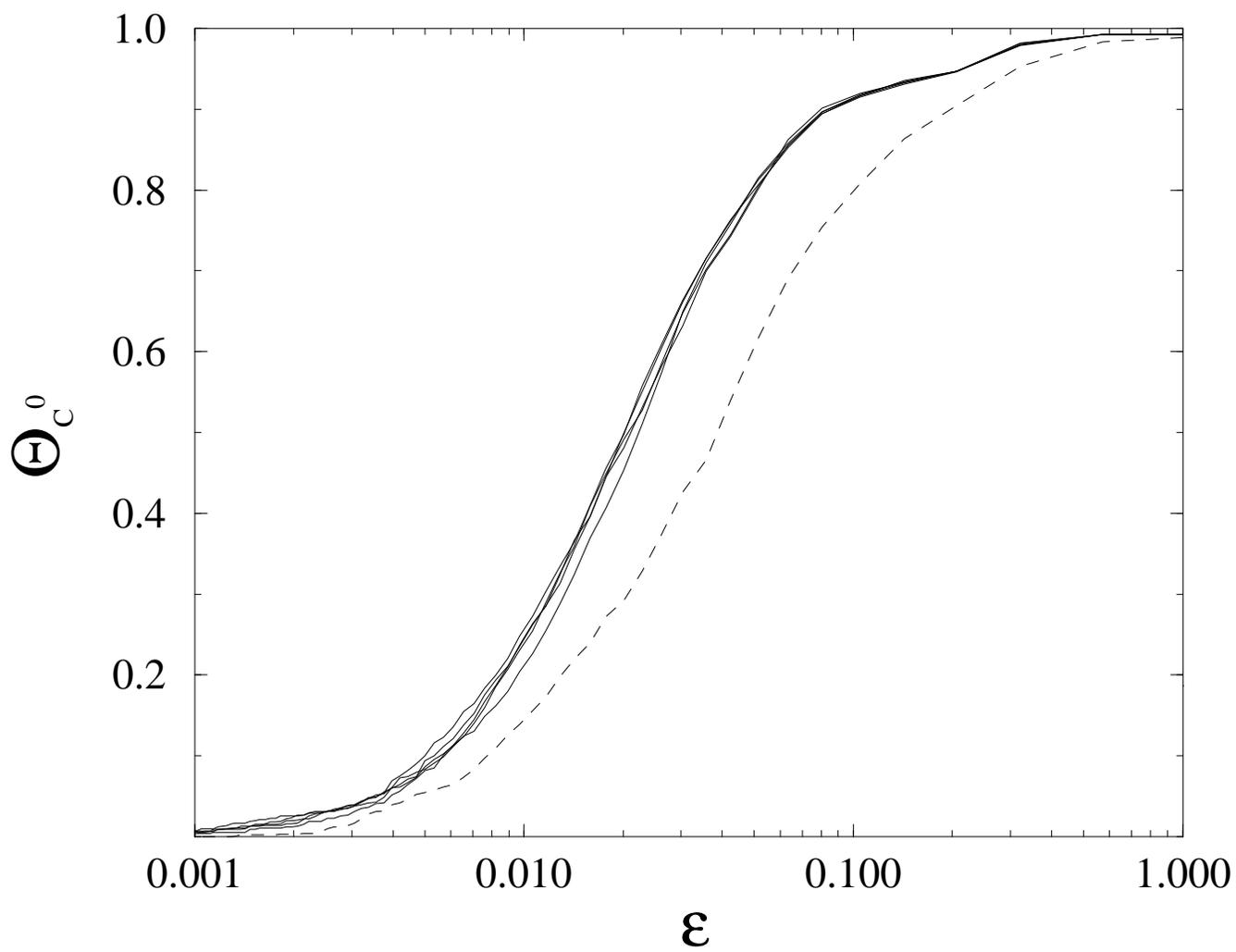
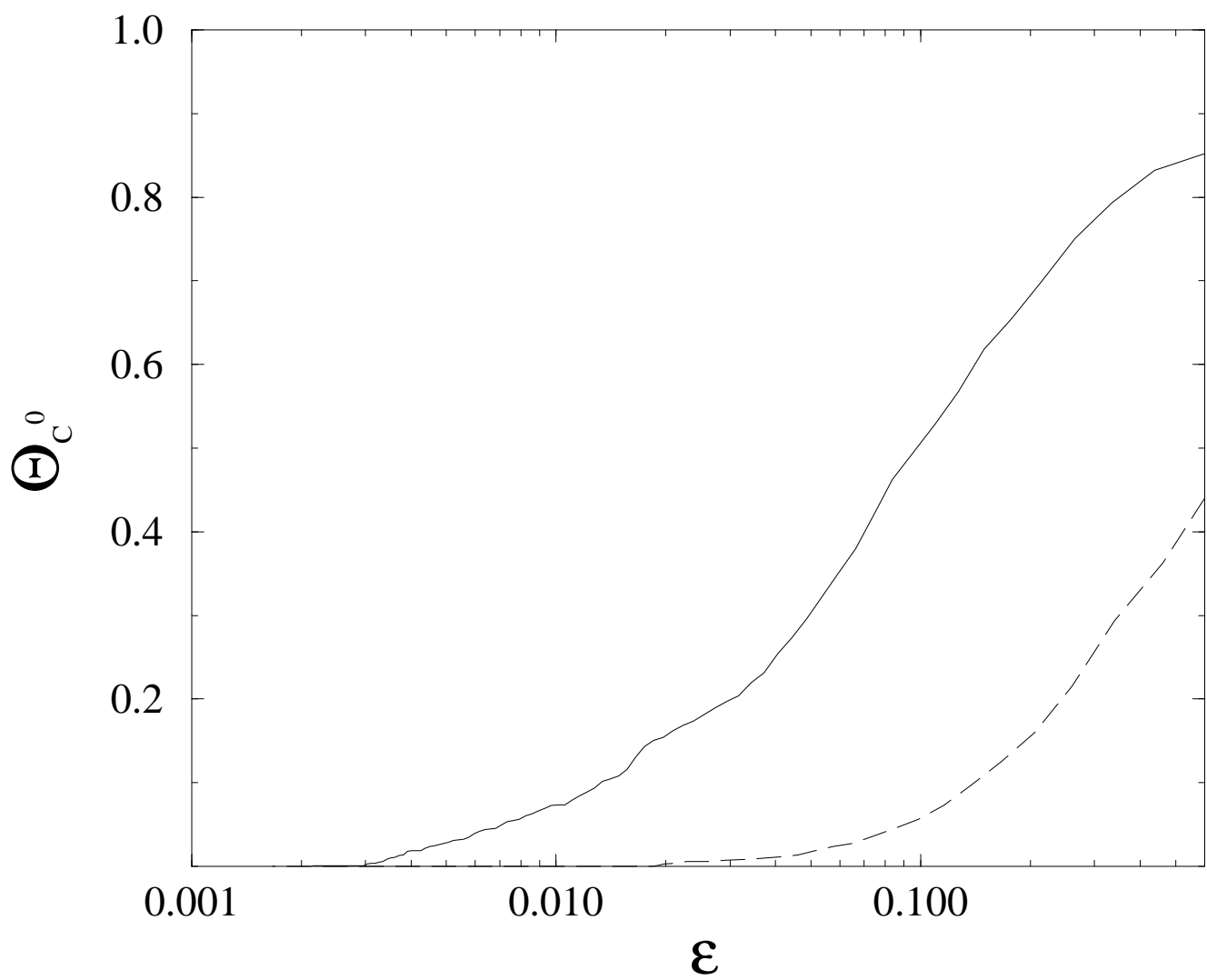
\end{document}